\documentstyle[12pt,epsf]{article}

\begin{document}

\begin{center}{{\bf
DECONFINEMENT IN QCD WITH DYNAMICAL QUARKS }
\footnote{Supported by Bundesministerium 
f\"ur Wissenschaft, Forschung und Kunst of Austria}  \\
\vglue 1.0cm
{O.~Borisenko, M.~Faber}\\
\vglue 0.2cm
\baselineskip=14pt
{\it Institut f\"ur Kernphysik,  Technische Universit\"at Wien,}\\
\baselineskip=14pt
{\it Wiedner Hauptstr. 8-10, A-1040 Vienna, Austria}\\
\vglue 0.4cm
{and G.~Zinovjev}\\
\vglue 0.2cm
\baselineskip=14pt
{\it Institute for Theoretical Physics, Ukrainian Academy of Science,
Kiev 143}\\
\vglue 1.4cm
{ABSTRACT}}
\end{center}
\vglue 0.6cm

We study the phase structure of full QCD within the canonical ensemble
(CE) with respect to triality in a lattice formulation. The procedure
for the calculation of the effective potentials (EP) in the CE is given.
We calculate the EP for the three dimensional $SU(2)$ gauge model at finite
temperatures in the strong coupling region. The potential exhibits a genuine
 deconfinement phase transition unlike the similar potential obtained
in the grand canonical ensemble (GCE) which demonstrates explicit $Z(N_c)$
symmetry breaking at any temperature. Furthermore, we investigate
the EP with the chiral condensate included. In contradiction to other
authors, we find chiral symmetry restoration in all
 triality sectors. In the scheme with massless staggered fermions
we observe chiral symmetry restoration accompanying a deconfinement
phase transition of first order. Above the critical point, besides
two $Z(2)$ symmetric  "deconfining" vacua there exists a metastable
"confining" vacuum in a wide region of the $(N_t,\gamma)$-plane. Such 
a picture could be interpreted as an indication on a mixed state 
of hadrons and quarks in the vicinity of the critical line.

\newpage

A comprehensive understanding of the phase structure of Quantum Chromodynamics
(QCD) is the basis for many phenomenological applications,
for possible signals of the quark-gluon plasma and other topics 
(see, for instance, the review on the Symposium 'Lattice-94' \cite{detar}).
In numerical calculations a number of new results has been obtained.
Nevertheless, the theoretical
understanding of the phase structure of full QCD, despite huge efforts,
is almost on the same level as it has been around fifteen years ago.
In our view, one of the central points of the problem is a reliable theoretical
determination of the deconfinement phase transition in QCD with dynamical fermions.
It was understood in the beginning of eighties \cite{gl} that in pure
gluodynamics deconfinement can be described in terms of the spontaneous
breakdown of the global $Z(N_c)$ symmetry of the vacuum. An appropriate
order parameter, the Polyakov loop,  was proposed to distinguish
the confined phase of static colour charges from the deconfinement.
Since loops of dynamical quarks violate in general the $Z(N_c)$ symmetry explicitly,
a naive application of this picture to full QCD is not possible.
At large values of quark masses a reminiscent
of deconfinement can be seen in the behaviour of some thermodynamical
functions. In QCD with massless quarks the behaviour of the quark condensate
demonstrates a phase transition associated with the restoration of chiral
symmetry. It was stressed, however, that
``with two flavors of light quarks  in the staggered fermion scheme,
there appears to be a dramatic crossover, but so far no evidence
for a genuine phase transition'' \cite{detar} (see also \cite{smilga}
for review of the thermal transition in QCD).

Recently, in several articles \cite{preprint,trl,versus} strong arguments have
been given that the general statement ``dynamical quarks in QCD breaks
$Z(N_c)$ symmetry explicitly'', is not valid.   It has been discussed 
that the fermionic
determinant in finite temperature QCD describes a grand canonical ensemble
(GCE). One can restrict the statistical averaging procedure to a canonical
ensemble (CE) with $n$-ality zero which may still be the grand canonical with
respect to baryon number. This CE does not break $Z(N_c)$ symmetry explicitly.

In Ref.\cite{trl} we speculated about the possible QCD phase structure.
We gave some arguments that there should be a genuine deconfinement phase
transition while in the GCE only a crossover has been observed.
Here, we would like to supplement our intuitive arguments by
a calculation of the effective potential in the CE at strong coupling.
Since the $Z(N_c)$ symmetry is not explicitly broken in this ensemble by
dynamical quarks, one may hope to find a phase transition similarly to pure
gluodynamics. Furthermore, we calculate the effective potential with the chiral
condensate included.
We shall show that the chiral symmetry is restored in all $n$-ality sectors.
This is in disagreement with the GCE description of full QCD 
\cite{chr,stp} and is another argument in favour of our approach.
First,  we review our general strategy and
then apply it to (2+1)-dimensional QCD with Wilson fermions.
Further,  we use our approach to study the behaviour of chiral symmetry
within the CE in the scheme with staggered fermions.

Let us consider the effective model of the Polyakov loops (PL) $W_{\vec{x}}$ in
the strong coupling region calculated first in Ref.\cite{pot}:
\begin{equation}
\int dU_{n}(x) e^{- S_G} \approx
e^{\gamma \sum_{\vec{x},n} Sp W_{\vec{x}} Sp W_{\vec{x}+n}},
\label{1}
\end{equation}
\noindent
where $S_G$ is the pure gluonic action and $\gamma \sim (\frac{1}{g^2})^{N_t}$.
For a system with heavy quarks one can apply the hopping parameter
expansion of the fermionic determinant in the scheme with Wilson fermions
\begin{equation}
Z_q = Det[ D_{xy} ]  \approx
e^{h \sum_{\vec{x}} Sp W_{\vec{x}} + O(h^2)},
\label{2}
\end{equation}
\noindent
with the hopping parameter $h = (\frac{1}{2d+m})^{N_t}$.
Combining (\ref{1}) and (\ref{2}) one gets for the effective theory of PLs
in the GCE with respect to triality
\begin{equation}
Z_{GCE} = \int \prod_{\vec{x}} dW_{\vec{x}}
\exp [\gamma \sum_{\vec{x},n} Sp W_{\vec{x}} Sp W_{\vec{x}+n}
+ h \sum_{\vec{x}} Sp W_{\vec{x}}].
\label{3}
\end{equation}
\noindent
It is well known that the model (\ref{3}) does not demonstrate a critical
behaviour at any $\gamma$ and $h$. This is a consequence of the explicit
violation of the global $Z(N)$ symmetry in the GCE.

Generalizing a simple technics for describing the system of zero triality
\cite{trl,versus}, one obtains the corresponding partition function in the CE
\begin{equation}
Z_{CE} = \frac{1}{N_c} \sum_{k=1}^{N_c} \int \prod_{\vec{x}} d L_{\vec{x}}
\exp [\gamma \sum_{\vec{x},n} L_{\vec{x}} L_{\vec{x}+n}
+ h \sum_{\vec{x}} z_k L_{\vec{x}}],
\label{4}
\end{equation}
\noindent
where $L_{\vec{x}} = Sp W_{\vec{x}}$ is the trace of the Polyakov loop 
matrix and $z_k = \exp [\frac{2\pi i}{N_c} k]$ fixes the phase sector. 
The integration in (\ref{4}) has to be
done with the invariant measure for the $SU(N)$ group.
We discuss now the $SU(2)$ gauge group.
A naive effective potential for the PL in the CE for $N_c=2$
is of the form \cite{versus}
\begin{equation}
V_{eff} = 2d\gamma {L^2} + \ln (1-L^2) + \frac{1}{N} \ln (\cosh NhL),
\label{5}
\end{equation}
\noindent
where $N$ is the number of lattice sites at a given time-slice.
In the thermodynamical limit we have the term $\mid Lh \mid$ in (\ref{5})
which breaks $Z(2)$ symmetry at any temperature. It follows
that $<L>$ always differs from zero in this approximation. One could
interpret this breaking as dynamical caused by the quark sea
at low temperatures and as spontaneous at high temperatures \cite{versus}.
Thus, if this is the case we have two different mechanisms of breaking.
To see the critical behaviour we would need an order parameter (OP)
which could distinguish between these mechanisms. The PL obviously
is not suitable for this goal.

The starting point of our calculations was the simple observation that at least
in the zero temperature case (or in the limit of infinite coupling) the naive
mean-field does not lead to the expected result: $Z(N_c)$ should not
be broken and CE and GCE descriptions have to coincide since in fact
we have no ensemble at zero temperature but only a system in its ground state:
The expectation value of the PL at zero temperature equals zero,
whereas the naive effective potential as well as the GCE predicts
$Z(N_c)$ is broken and the PL differs from zero in this limit.

In this respect the essential point is:

In different phase sectors in (\ref{4}) the minimum of the effective 
action is reached for different $Z(N_c)$ configurations.
This is very important in the disordered confinement phase where 
the PL may take any value. When we consider the different
phase sectors on the same configurations it follows that
only one of them will survive in the thermodynamical limit.
However, if these configurations are indeed disordered, averaging over them
should give the opposite result: reaching equal minima but for different
configurations, two sectors contribute to the thermodynamical limit
in the confinement phase, cancelling all noninvariant contributions.

Hence, the main idea is to execute the summation over $Z(N)$ configurations
contained in the PL and only then to apply the mean field approximation.
Doing in this way we can reproduce the true minima of the effective
potential in each triality sector. To accomplish this one uses the formula
\begin{equation}
\int dU G(U) = \frac{1}{N_c}\sum_z \int dU G(zU),  \  z \in Z(N_c),
\label{6}
\end{equation}
\noindent
which is valid for any compact group and its central subgroup $Z(N_c)$.
This formula gives us a chance to verify whether the final result depends
explicitly on the $Z(N_c)$ configurations contained in the initial action
$S_G$ or not.  If yes, we have the $Z(N_c)$ symmetry broken at all temperatures.
In the other case, we can find a spontaneous breaking only at high temperatures.

Thus, we start from the following general expression for the EP
applying Eq.(\ref{6})
\begin{eqnarray}
e^{N[V_{eff}(L) - V(T=0)]} = \frac{1}{N_c} \sum_{k=1}^{N_c}
\sum_{ \{s_x\},s_x \in Z(N_c)}
\int D\mu (L_x) \exp [S_{eff}(z_k s_x L_x)] \nonumber   \\
\delta \left [ \sum_x s_x L_x - NL \right ],
\label{7}
\end{eqnarray}
\noindent
where we 0 the contribution to $V_{eff}$ at zero temperature.
We choose the gauge $U_0(x,t)=1$ for $t \in \{1,2,..,N_t-1\}$ and
$U_0(x,N_t)=W_x$ with diagonal $W_x$.

The effective action $S_{eff}$ in (\ref{7}) is defined as an integral over
all space components of gauge fields with the Wilson fermionic determinant 
included.
Its calculation is clearly far from our present ability. Using the strong
coupling and the hopping parameter expansions as well as the mean field
approximation one arrives to the following effective potential in the CE
\begin{eqnarray}
\exp [NV_{eff}^{CE}] =
\frac{1}{N_c} \sum_{k=1}^{N_c} \ \ \sum_{ \{s_x\}, \ s_x \in Z(N_c)}
 \nonumber   \\
\exp [\gamma L^2\sum_{x,n} Re s_xs_{x+n}^{\star}
+ h L\sum_x z_k Re s_x + V_{IM} ],
\label{8}
\end{eqnarray}
\noindent
with $\gamma$ and $h$ defined in (\ref{1}) and (\ref{2}), respectively.
$V_{IM}$ stays for the invariant measure for the $SU(N)$ group.
The summation over the $N_c$ n-ality sectors guarantees that only
states contribute to the effective potential whose number
of spins $s_x$ is a multiple of $N_c$.

Eqs.(\ref{7})-(\ref{8}) are still valid in arbitrary dimensions and
for any $SU(N)$ group.
To make some further progress we consider in the following the $SU(2)$ gauge 
theory in
$d=2+1$ dimensions. Then, our effective model (\ref{8}) is two dimensional.
We have to solve the $2-d$ Ising model with an effective coupling
in an external field and to investigate the effective potential obtained.

After some calculations we find  the following bound for the EP
\begin{eqnarray}
V_{eff}^{CE}(L) \ \leq \  V_{IM}(L) + \frac{1}{N}\ln Z_I(\gamma_L) +
\ln \cosh Lh +  \nonumber   \\
\frac{1}{N} \ln \frac{1}{N}\sum_{r\neq 0}\cosh [N\tanh Lh(<s(0)s(r)>)^{1/2}].
\label{9}
\end{eqnarray}
\noindent
$Z_{I}$ in (\ref{9}) is the partition function of the $2-d$ Ising model
without external magnetic field
calculated at the effective coupling $\gamma_L = \gamma L^2$.

We used the following inequality to calculate the sum over
pair correlation functions\footnote{In our terms the effective potential
presented in \cite{versus} would correspond to $\Gamma \leq N^{2k}$. It is
obvious, however, that the bound (\ref{10}) is lower.}
\begin{equation}
\Gamma = \sum_{x_1\neq x_2\neq ... \neq x_{2k}} <\prod_{i=1}^{2k}s(x_i)> \
\leq  \  N^{2k-1}\sum_{r\neq 0}<s(0)s(r)>^k.
\label{10}
\end{equation}
\noindent
The EP in (\ref{9}) exhibits a very essential feature of the CE description,
namely all odd correlation functions are identically cancelled due to 
the global
summation over the center elements $z_k$. Just these contributions lead to the
explicit violation of the $Z(2)$ symmetry in the GCE description. It should be
stressed that the cancellation happens exactly and is not a consequence of
the approximations used.

To make the following arguments as transparent as possible and to
promote the analytical evaluations to the very end we use simple asymptotics
for the $2-d$ Ising model with an external field at small and large values
of the effective coupling $\gamma_L$ \cite{baxter}. This leads up to a constant
independent of $L$ to asymptotic expressions for the EP of the form 
\begin{eqnarray}
V_{eff}^{CE}(L,\gamma_L \ll 1) = \ln (1-L^2) + \ln \cosh Lh +
2\gamma L^2 (\tanh Lh)^2 +    \nonumber   \\
\gamma^2 L^4 [1 + 6(\tanh Lh)^2 - 7(\tanh Lh)^4] + O(\gamma^3)
\label{pot1}
\end{eqnarray}
\noindent
and
\begin{equation}
V_{eff}^{CE}(L,\gamma_L \gg 1) = \ln (1-L^2) + F(h).
\label{pot2}
\end{equation}
\noindent
The function $F(h)$ is defined as
\begin{eqnarray}
F(h) = 2\gamma L^2 + Xh + u^2\exp (-2Xh) + 2u^3\exp (-4Xh) +  \nonumber  \\
u^4 [-\exp (-4Xh) + 6\exp (-6Xh) + \exp (-8Xh)] + O(u^5),
\label{fh}
\end{eqnarray}
\noindent
where $u=\exp (-4\gamma L^2)$ and $X = \mid L \mid$.

Analyzing the EP (\ref{pot1}) one concludes that at sufficiently small
$\gamma$ and $h$ the maximum of the EP is located at $L=0$.
As $\gamma$ and/or $h$ increases two maxima with
$L = \pm \mid L_{max} \mid \neq 0$ appear dominating the EP. In this
approximation it is easy to find a critical line in the $(\gamma -h)$-plane.
We omit this simple calculation as further we study in details the more
interesting case with the quark condensate included.
The EP for $\gamma_L \gg 1$ (\ref{pot2}) is also
formally symmetric under $Z(2)$ transformations as it should be for
the CE at any temperature. However, because of the presence of the term
$Xh$ in (\ref{fh}) the EP develops two maxima at $L \neq 0$ for any $h$.
Thus, $Z(2)$ symmetry can be spontaneously broken at high temperatures
indicating the deconfinement phase transition. Since we have two
degenerate maxima for $\gamma \gg 1$, domain walls might separate
regions with positive and negative values of $L$, similar to pure
gluodynamics. Concerning this question see, however, \cite{kis}. 
It is worth to mention that the form of the EP at large
$\gamma_L$ is close both to the EP in the GCE and to the EP in the CE 
calculated in \cite{versus}. This means that the EP from \cite{versus} 
is only valid in the deconfinement phase.

Now, we would like to construct and to study an effective potential
for Polyakov loops and for the quark condensate $\sigma$ in the CE.
It has been argued in \cite{chr} that the chiral symmetry is restored
only in one n-ality sector (e.g., in the real phase for $SU(3)$) while
it is still broken in other sectors above the critical temperature where
the chiral condensate becomes zero in the real phase. Moreover,
in the phase arg$L = \pi$ of $SU(2)$ gauge theory the chiral condensate
is 1 for all temperatures \cite{stp}.
We shall show that this may only be valid in the GCE. Treating QCD within
the CE leads to chiral symmetry restoration in all triality sectors.

In the limit of vanishing quark mass the action for staggered fermions
is chiral symmetric. Therefore, we turn now to the scheme with staggered
fermions. For the effective action which includes effects of dynamical
quarks in terms of mesonic field $\bar{\Psi}_x\Psi_x = \sigma_x$
we use the effective action of Ref.\cite{og}. 0 this
action to the CE, we get
\begin{eqnarray}
\exp [NV_{eff}^{CE}(L, \sigma)] = \frac{1}{2} \sum_{k=1}^2 \sum_{ \{s_x\},s_x \in Z(N)}
\int_{-\infty}^{\infty}\prod_x d\sigma_x
\exp [\gamma L^2\sum_{x,n} s_xs_{x+n} + V_{IM}]  \nonumber  \\
\exp [- N_t \sum_{x,y}\sigma_x V_{xy}^{-1}\sigma_y ]
\prod_x \left [ \cosh N_tE + z_k L s_x  \right ]
\prod_x \delta \left [ \sum_x \sigma_x - N\sigma \right ],
\label{14}
\end{eqnarray}
\noindent
where (for $d=2$)
$$
\sinh E = m + \sigma_x
$$
and
$$
 V_{xy} = \frac{1}{4} \sum_{n=1}^d (\delta_{y,x+n} + \delta_{y,x-n}).
$$

To get a relatively simple form for the EP we use the simplest mean-field
approximation for mesonic fields $\sigma_x$ neglecting all fluctuations.
Taking again the asymptotics for the two dimensional Ising model as
described above we end up with the following expressions for the EP
\begin{eqnarray}
V_{eff}^{CE}(L, \sigma) = \ln (1-L^2) + \ln H - N_t \sigma^2 +
\ln \cosh (L/H) +  \nonumber   \\
2\gamma L^2 (\tanh L/H)^2 +
\gamma^2 L^4 [1 + 6(\tanh L/H)^2 - 7(\tanh L/H)^4] + O(\gamma^3)
\label{15}
\end{eqnarray}
\noindent
for $\gamma L^2 \ll 1$ and
\begin{equation}
V_{eff}^{CE}(L, \sigma) = \ln (1-L^2) + \ln H - N_t \sigma^2 + 2\gamma L^2 +
F_c(H)
\label{16}
\end{equation}
\noindent
for $\gamma L^2 \gg 1$. The function $F_c(H)$ is of the form
\begin{eqnarray}
F(H) = X /H + u^2\exp (-2X /H) 
+ 2u^3\exp (-4X /H) +   \nonumber   \\
u^4 [-\exp (-4X /H) + 6\exp (-6X /H) + 
\exp (-8X /H)] + O(u^5),
\label{fhc}
\end{eqnarray}
\noindent
where functions $u$ and $X$ were defined after Eq.(\ref{fh}).
We have denoted here
$$
H = \cosh N_tE.
$$
These formulae are valid in the region $X /H \ll 1$. In what 
follows we consider $\gamma$ and $N_t$ as independent variables and study EPs
for different values of $\gamma$ and $N_t$. First of all, we observe that
the EP in (\ref{15}) for sufficiently small $\gamma$ has the maximum at   
$L=0$. The EP in (\ref{16}) exhibits
$Z(2)$ symmetry breaking in the proper region of the convergence of the
series in (\ref{fhc}) although, similar to the EP in (\ref{pot2}), it
is also center symmetric. Hence, with respect to $Z(2)$ symmetry the
behaviour of the system is close to the previous case with Wilson fermions.

\begin{figure}[htb]
\centerline{\epsfxsize=9cm \epsfbox{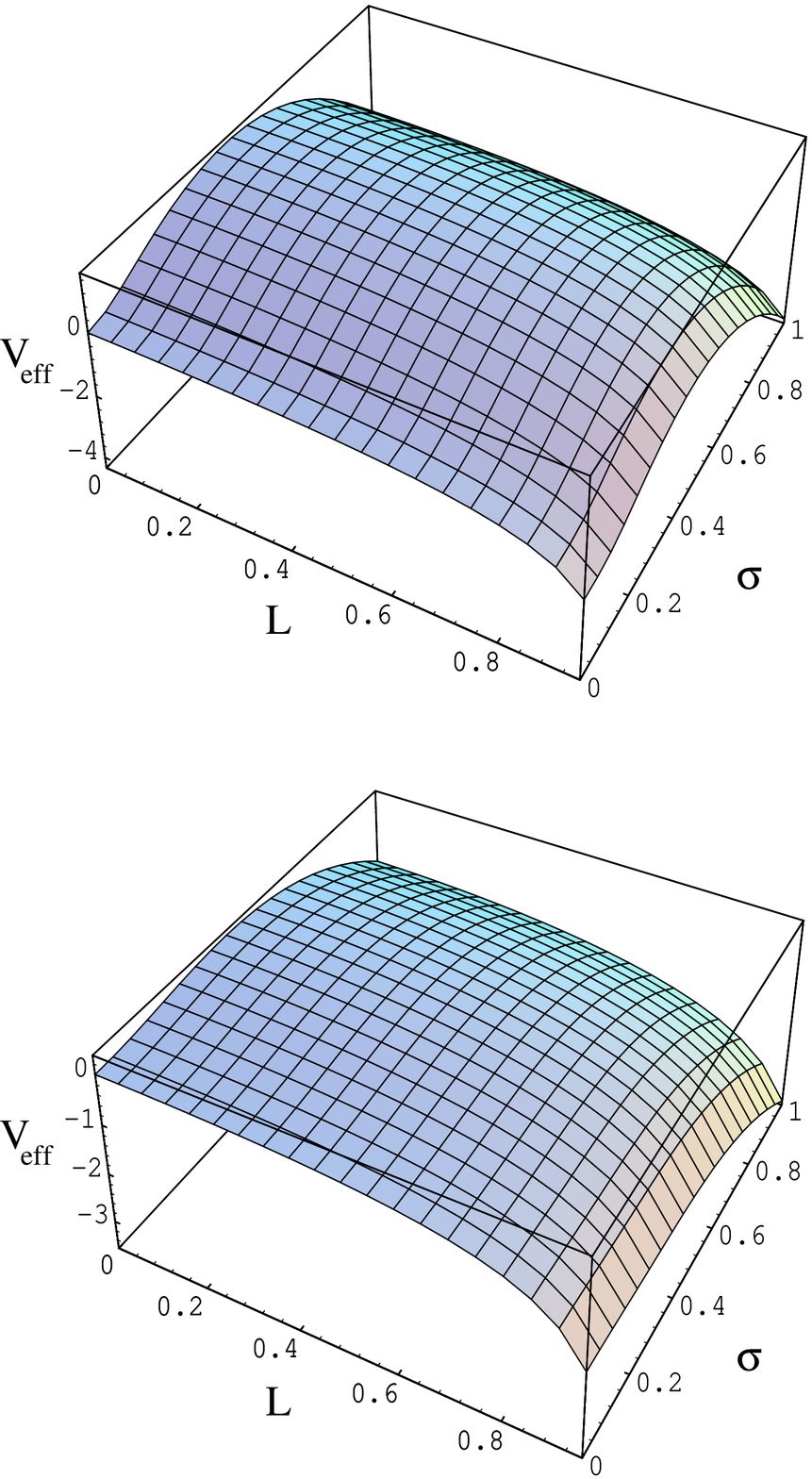}}
\caption{Effective potential in the confinement 
region for $\gamma =0.1$, $N_t=10$ above and $N_t=4$ below.}
\label{eff_pot}
\end{figure}

\begin{figure}[t]
\centerline{\epsfxsize=9cm \epsfbox{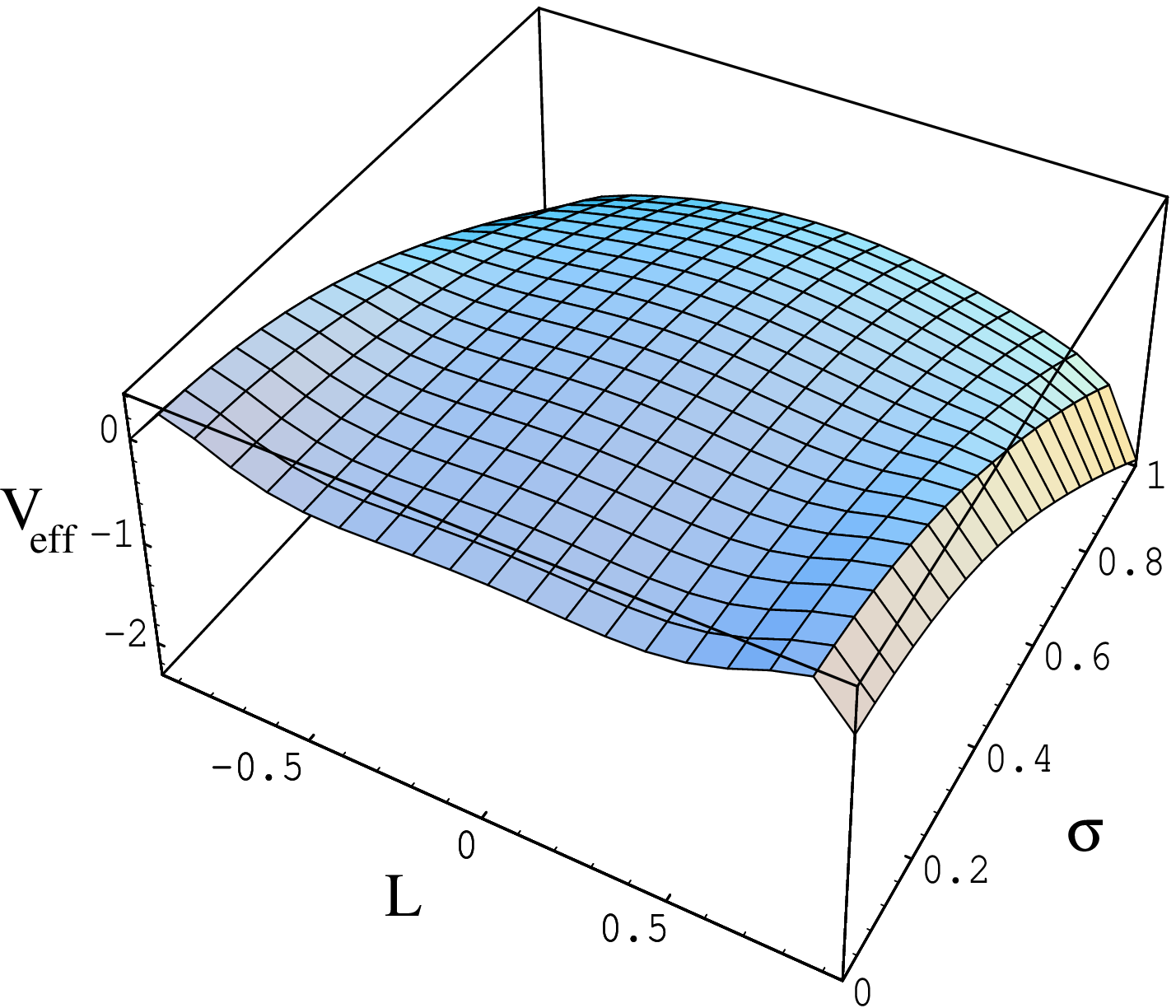}}
\caption{Effective potential around the critical line
for $N_t=2$ and $\gamma =0.8$.}
\label{fig2}
\end{figure}

We suppose now that the EP (\ref{15}) reflects all
qualitative features of the theory in the region $\gamma \leq 1$.
This may be the case in the vicinity of $L \approx 0$ since the real
expansion parameter is $\gamma L^2$ and therefore the series in
Eq.(\ref{15}) can still converge. To study the chiral symmetry behaviour
we have plotted the EP (\ref{15}) for several values of $N_t$ and
$\gamma$ taking the quark mass $m=0$.
Fig. \ref{eff_pot}  shows the EP at $\gamma = 0.1$ for $N_t=10$ and for $N_t=4$. 
Because these graphics are obviously symmetric with 
respect to $Z(2)$ transformations we plotted them for positive $L$ only.
They demonstrate that the maximum of the EP is located at $L=0$.
One sees that at large $N_t$ the chiral symmetry is strongly broken 
($\sigma \neq 0$). As $N_t$ decreases (temperature increases) the chiral
symmetry tends to restore and $\sigma \rightarrow 0$.
Fig. \ref{fig2} presents the EP (\ref{15}) in the vicinity of the critical line
for $N_t=2$ and $\gamma = 0.8$. One observes the main maximum at
$L \neq 0$ and $\sigma = 0$ for these values of the parameters.

\begin{figure}[t]
\centerline{\epsfxsize=9cm \epsfbox{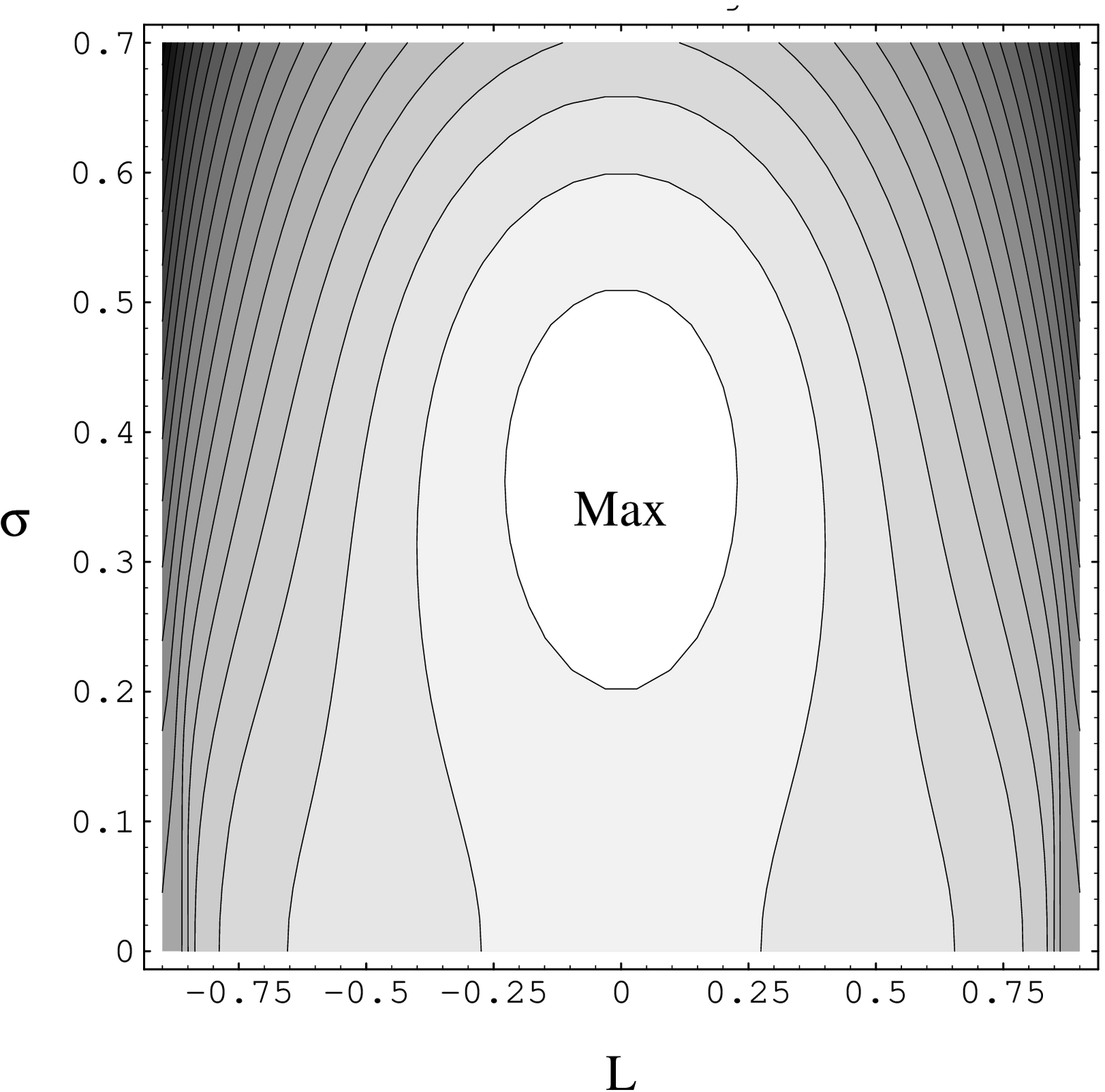}}
\caption{Contour Plot of the effective potential in 
the confinement region for $N_t=3$ and $\gamma =0.5$.}
\label{fig3}
\end{figure}

\begin{figure}[t]
\centerline{\epsfxsize=9cm \epsfbox{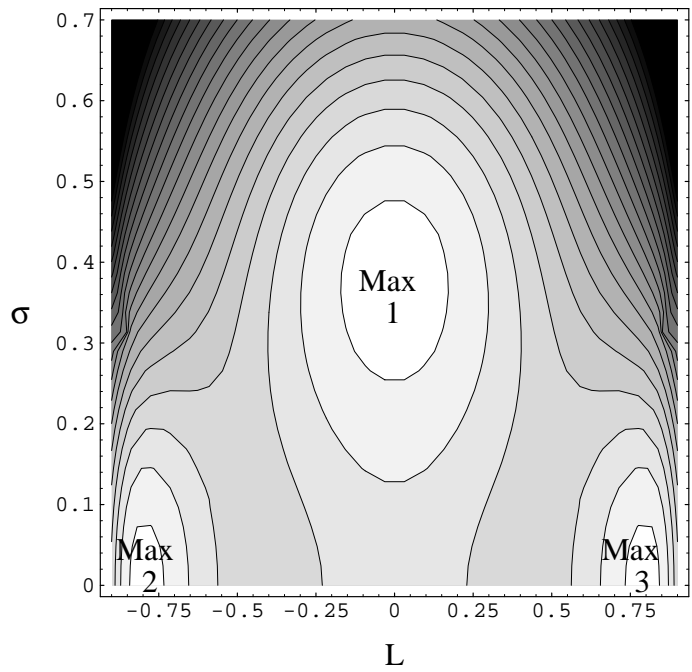}}
\caption{Contour Plot of the effective potential at
the critical point $\gamma_c =0.68$ for $N_t=3$.}
\label{fig4}
\end{figure}

\begin{figure}[t]
\centerline{\epsfxsize=9cm \epsfbox{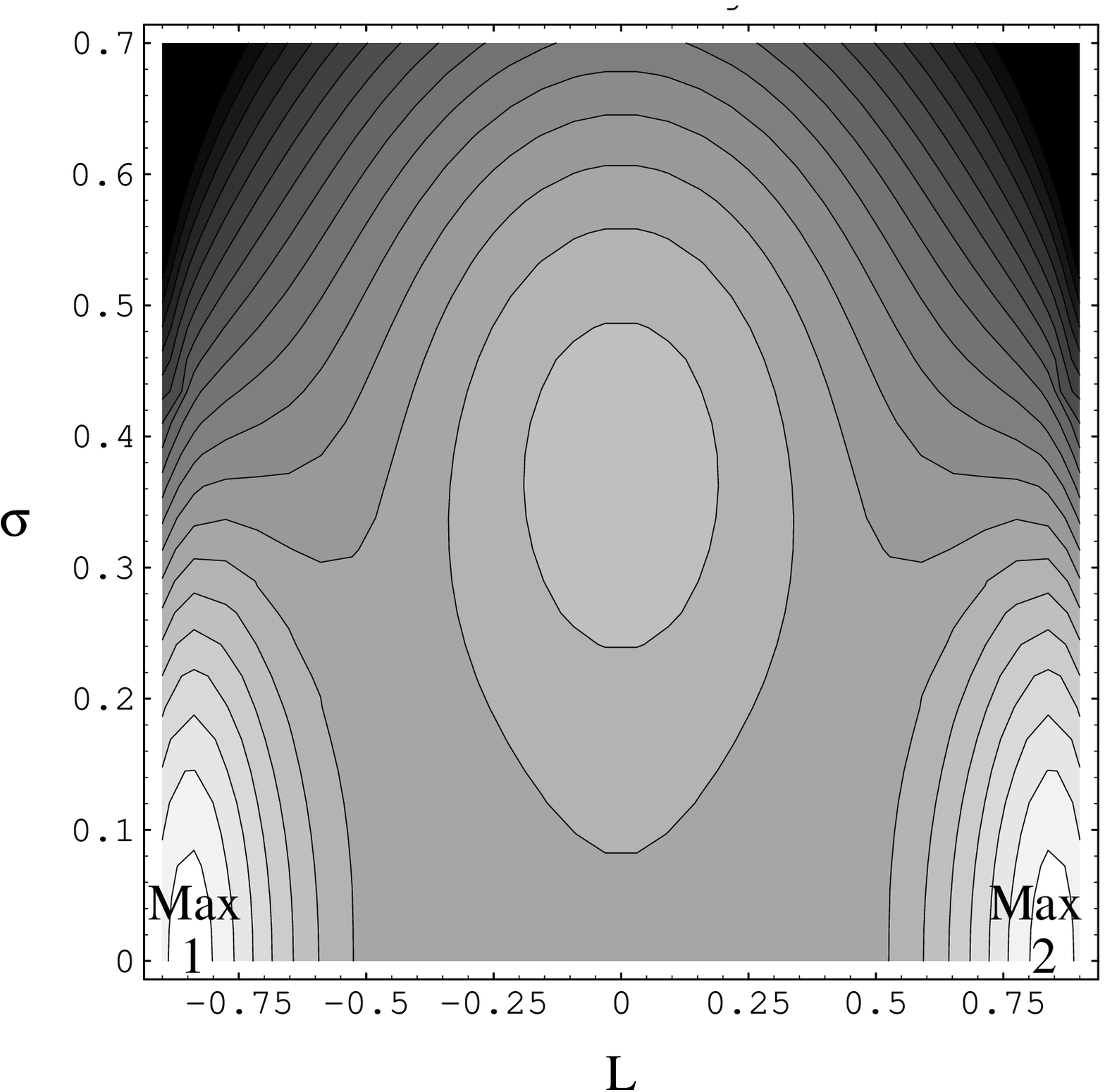}}
\caption{Contour Plot of the effective potential in 
the chiral symmetric phase for $N_t=3$ and $\gamma =0.68$.}
\label{fig5}
\end{figure}

In all cases for $N_t \in [1-6]$ and $\gamma \in [0.5-0.9]$
the spontaneous breaking of $Z(2)$ symmetry is accompanied by
chiral symmetry restoration since $\sigma =0$ above the critical point.
The phase transition is of first order.
The important fact is that $\sigma =0$ in both sectors, i.e. for positive
and negative values of the PL. Therefore, the chiral symmetry is restored
in both sectors. This is in contradiction with the GCE description
where only one phase exhibits such a restoration \cite{chr,stp}.
As $\gamma$ grows, the symmetric maxima of the EP are smoothly moving
from each other to the values $L=\pm 1$ for the PL. In our approximation
it is impossible, however to establish reliably whether they can reach
these values because in this region we are far away from
the strong coupling where our EP can be trusted.
We have shown this behaviour in Figs.~\ref{fig3}--\ref{fig5} 
plotting contour graphics for 
$N_t=3$. Fig. \ref{fig3} presents the confinement region with the only maximum 
at $L=0$. Around $\gamma \approx 0.6$ two symmetric deconfining maxima
with $\sigma = 0$ start to develop. Fig. \ref{fig4} shows the critical point
$\gamma = 0.68$ where confining and deconfining maxima become degenerate.
Fig. \ref{fig5} shows the deconfinement phase where the maximum with 
$L=0$ is metastable.    
There exists a rather wide region in the $(N_t,\gamma)$-plane where we 
find the coexistence of the confining vacuum with $\sigma \neq 0$ 
and the two degenerate deconfining vacua with $\sigma = 0$. One of them 
is always metastable but physically acceptable because they have normal 
physical features in contrary to the so-called $Z(N)$ metastable phase.
This picture implies the existence of different "confining" and 
"deconfining" bubbles around the critical line with nonzero surface tension. 
One could interpret
this vacuum structure as an indication of a mixed state in hot nuclear
matter which has been discussed for more than ten years as a hyphotetical
state in the phenomenology of the quark-gluon plasma but the existence of 
which, to our knowledge, has not been confirmed up to now 
from first principles.

Let us briefly summarize our results and discuss some open questions.

In this letter we gave a prescription for the calculation of the EP
in the CE with respect to triality. We 1 that all our results
are valid in $d=3+1$ dimensions for the $SU(3)$ gauge group though
the analytical calculations in this case are more difficult.
The EP in the CE demonstrates reacher phase structure in contrast 
to the corresponding potential in the GCE. Three main points 
distinguish our approach from the conventional one: 1) The deconfinement 
phase transition manifests itself in the CE similarly to pure 
gluodynamics. 2) The chiral symmetry is restored together with spontaneous
breaking of $Z(N_c)$ symmetry and this restoration takes place in all
$n$-ality sectors. 3) A reach vacuum structure around the critical line
indicates the existence of a mixed state in hot nuclear matter.

We showed these properties both for Wilson and for staggered fermions. The CE 
exhibits closer interrelation between deconfinement and chiral symmetry
restoration. This connection might be even tighter if we had not used
the approximation $L/H \ll 1$. It could be very instructive in this respect 
to include a nonzero baryonic number in the present approach as well as 
a finite value for the quark mass.

It has been argued in \cite{trl,sem} that $Z(N_c)$ global symmetry at finite
temperature has a physical interpretation in terms of the charge of physical 
states. In our case this is the triality charge of fundamental quarks.
The following question can be posed at this point: 
is the $Z(2)$ symmetry really broken and $L \neq 0$ above the critical point
or does the presented picture give only a formal way to discover
a critical behaviour? The inconsistency may arise, since if the $Z(N_c)$
symmetry is broken, there could be non-zero triality states contributing to 
the partition function \cite{sem}.  
It would then intuitively seem that the projection onto
zero triality states and the spontaneous breakdown of $Z(N_c)$ symmetry
are in contradiction with each other. The possible answer might be 
that the CE with respect to triality and projection of the physical
observables onto zero triality states are different things:
in the former case, if $Z(N_c)$ is broken, there are nonzero triality
contributions to the free energy at high temperatures, whereas in the 
later case such contributions are absent. However, we tend to 
the opinion that in the case of the CE description of finite 
temperature QCD non-zero triality contributions are all vanishing in
the thermodynamical limit. That such a vanishing is not in contradiction
with the Debye screening in the deconfinement phase has been discussed
in \cite{preprint}. An interesting approach is to consider the Hamiltonian
formulation and to interpret all physics exclusively in terms of 
physical space variables \cite{kis}. In this respect we reckon, that 
the application of the consideration of Ref.\cite{kis} to the theory 
with dynamical fermions described by the CE with respect to triality 
would give interesting results concerning critical behaviour of full 
QCD. We do not, however, think that $Z(N_c)$ symmetry is unphysical
as has been discussed in \cite{kis} and \cite{smilga1}. It is really so, 
that $Z(N_c)$ symmetry acts as an identity on all physical states. 
The reason for this is $Z(N_c)$ global transformations of zero
temperature QCD together with the substitution 
$$
\Psi (x,\beta) \rightarrow \exp(i\frac{2\pi}{N}k)\Psi (x,0)
$$     
\noindent
for the fermionic fields $\Psi$
determines the triality of quark states at finite temperature \cite{trl}.
In this sense, the $Z(N_c)$ symmetry is a physical symmetry
(i.e., if triality charges are physical charges). One has to be
0, nevertheless, treating this symmetry at finite temperature since
it may occur that not only states having zero triality charge contribute
to the partition function.  
The CE ensures that only states contribute to the partition function
which transform identically under above transformations.

\vglue 0.6cm

\end{document}